\documentclass[prd,aps,showpacs,onecolumn]{revtex4}
\usepackage{amsmath}
\usepackage{amssymb}
\usepackage{color}
\usepackage{cancel}
\usepackage{bbm}
\usepackage{bm}
\usepackage{graphicx}
\usepackage{maybemath}

\newcommand{\dd}{\mathrm d}
\newcommand{\ii}{\mathrm i}
\newcommand{\ee}{\mathrm e}
\newcommand{\calC}{\mathcal C}
\newcommand{\calP}{\mathcal P}
\newcommand{\calT}{\mathcal T}
\newcommand{\calL}{\mathcal L}
\newcommand{\calU}{\mathcal U}
\newcommand{\calV}{\mathcal V}
\newcommand{\eV}{\mathrm{eV}}
\newcommand{\keV}{\mathrm{keV}}
\newcommand{\MeV}{\mathrm{MeV}}
\newcommand{\GeV}{\mathrm{GeV}}

\newcommand{\m}{\mathrm{m}}

\newcommand{\ns}{\mathrm{ns}}

\newcommand{\plus}{{\mbox{{\bf{\tiny +}}}}}

\begin{document}

\bibliographystyle{myprsty}

\title{Tachyonic Field Theory and Neutrino Mass Running}

\author{U. D. Jentschura}

\affiliation{Department of Physics,
Missouri University of Science and Technology,
Rolla, Missouri 65409-0640, USA}

\affiliation{Institut f\"ur Theoretische Physik,
Universit\"{a}t Heidelberg,
Philosophenweg 16, 69120 Heidelberg, Germany}

\begin{abstract}
In this paper three things are done. (i)~We investigate the 
analogues of Cerenkov radiation for the decay of a superluminal
neutrino and calculate the Cerenkov angles for the 
emission of a photon through a $W$ loop, and for a collinear 
electron-positron pair, assuming the tachyonic dispersion 
relation for the superluminal neutrino. 
The decay rate of a freely propagating neutrino is found to
depend on the shape of the assumed dispersion relation,
and is found to decrease with decreasing tachyonic mass of the 
neutrino.
(ii)~We discuss a few properties of the tachyonic Dirac 
equation (symmetries and plane-wave solutions), which 
may be relevant for the description of superluminal 
neutrinos seen by the OPERA experiment, and discuss the calculation 
of the tachyonic propagator.
(iii)~In the absence of a commonly accepted tachyonic field theory,
and in view of an apparent ``running'' of the observed neutrino 
mass with the energy, we write down a model Lagrangian, 
which describes a Yukawa-type interaction of a 
neutrino coupling to a scalar background field via
a scalar-minus-pseudoscalar interaction.
This constitutes an extension of the standard model.
If the interaction is strong, then it leads to 
a substantial renormalization-group ``running'' of the 
neutrino mass and could potentially explain the 
experimental observations.
\end{abstract}


\maketitle

%
%
\section{Introduction}

For subluminal particles (``tardyons''),
the dispersion relation for the energy $E$ in terms of the velocity $v$ is given by 
$E = m/\sqrt{1-v^2}$ (with $v < 1$), 
and for superluminal particles (``tachyons''),
it reads as $E = m/\sqrt{v^2-1}$ with~$v > 1$.
Therefore, the ``light barrier'' at $v = 1$ (we set the speed of light 
equal to one) looks like an (infinitely) elevated mountain
in terms of the energy of a relativistic particle.
Recami~\cite{Re2009} quotes Sudarshan with
reference to an imaginary demographer who studies population patterns on 
the Indian subcontinent:
``Suppose a demographer calmly asserts that there are no people North of the
Himalayas, since none could climb over the mountain ranges! That would be an
absurd conclusion. People of central Asia are born there and live there: they
did not have to be born in India and cross the mountain range. So with
faster-than-light particles.''

In the early morning hours of 23 February 1987 (at 2'52'36''), 
an unexpected neutrino bunch arrived at the LSD detector under the Mont Blanc 
roughly 4.5 hours before the rest of the neutrinos from SN1987A,
and before the supernova became visible~\cite{DaEtAl1987}.
We are currently facing mounting evidence that 
neutrinos may be genuinely superluminal particles (``tachyons'').
The MINOS experiment~\cite{AdEtAl2007} has measured 
superluminal neutrino propagation velocities
which differ from the speed of light by a relative factor of 
$(5.1 \pm 2.9) \times 10^{-5}$ at an energy of 
about $E_\nu \approx 3 \, \GeV$,
supporting an earlier FERMILAB experiment 
where the trend of the data also pointed toward 
superluminal neutrinos~\cite{KaEtAl1979}.
This result has recently been confirmed by OPERA~\cite{OPERA2011v2} 
with better statistics and in a wider energy interval,
as detailed below.
One of the prime candidates for a genuinely
superluminal particle is the neutrino, which has
never been observed at rest.
A number of experimental groups have measured negative
mass squares for the electron neutrino from tritium beta decay 
endpoints~\cite{RoEtAl1991,AsEtAl1994,StDe1995,AsEtAl1996}
with mean values in the interval $-147 \,\eV^2 < m^2_{\nu} < 0$
for the electron neutrino mass square, at an energy of the order
of $E_\nu \approx 18 \, \keV$.
While some recent measurements indicate values consistent   
with a vanishing neutrino mass~\cite{WeEtAl1999,LoEtAl1999,BeEtAl2008}
at even lower energies, the mean value of the experimental data for  
$m^2_{\nu}$ (electron neutrino)
still is negative and of the order of a few negative $\eV^2$
(for an excellent overview, see Ref.~\cite{LABneutrino}).
The idea that neutrinos might be of tachyonic character
is not new~\cite{ChHaKo1985,Ch2000,Ch2002,JeWu2011jpa}.
Tachyonic neutrinos fulfill the dispersion relation
$E_\nu^2 - \vec p^{\,2} = -m_\nu^2$
with an (initially) constant parameter $m_\nu$.
The quantity $-m^2_\nu$ can be interpreted as the
negative mass square of the neutrino.
The current situation indicates the need for a 
convenient descriptions of tachyonic fermions.

Ever since the early days of relativity, the notion of superluminal propagation
has intrigued physicists~\cite{So1905}, and the name ``tachyon'' was eventually
coined in Ref.~\cite{Fe1967}.  The main problem in the description of a quantum
field theory with superluminal propagation is not the superluminal velocity
itself~\cite{BiDeSu1962}, but the construction of field operators and the time
ordering, which is in disarray because the time ordering of two space-time
points which are separated by a space-like interval is not invariant 
under (subluminal) Lorentz boosts.
Generally, it has been assumed that 
any particle in relativistic quantum theory should be 
described by a unitary irreducible representation of 
the Poincar\'{e} algebra or its supersymmetric
generalization. It may be necessary to relax this 
restriction somewhat in order to accommodate 
for a field theory of supersymmetric tachyons~\cite{BaSh1974,vDNgBi1985,XiJi1987}.
Three recent review articles~\cite{Re2009,Bi2009,Bo2009}
provide rather detailed background information on the
development of the theory of superluminal particles.

The recent OPERA experiment~\cite{OPERA2011v2} 
uses a baseline of $L = (731278.0 \pm 0.2) \, \m$.
Two clocks used in the measurement are 
accurately synchronized by a technique used 
to compare atomic clocks~\cite{DePe2003,Le2008}. 
It is of particular importance that the synchronization of the 
two systems was calibrated 
by the Federal Swiss Metrology
Institute METAS (Bundesamt f\"{u}r Metrologie) in 2008 
and verified in 2011 by the Federal 
German Metrology Institute PTB (Physikalisch-Technische
Bundesanstalt). As reported in Ref.~\cite{OPERA2011v2},
the difference between the time base of the CERN and OPERA 
receivers was measured to be $(2.3 \pm 0.9) \, \ns$ and is
taken into account in the evaluation of the measurement.
The four data bins are
\begin{subequations}
\label{eee}
\begin{align}
E_\nu =& \; 13.8 \, \GeV \,, \qquad
\delta t = (54.7 \pm 18.4 \pm 7.1) \, \ns \,, \qquad
\Delta = \frac{v-c}{c} = ( 2.24 \pm 0.75 \pm 0.29 ) \times 10^{-5} \,,
\\[0.77ex]
E_\nu =& \; 28.2 \, \GeV \,, \qquad
\delta t = (61.1 \pm 13.2 \pm 7.1) \, \ns \,, \qquad
\Delta = \frac{v-c}{c} = ( 2.50 \pm 0.54 \pm 0.29 ) \times 10^{-5} \,,
\\[0.77ex]
E_\nu =& \; 40.7 \, \GeV \,, \qquad
\delta t = (68.1 \pm 19.1 \pm 7.1) \, \ns \,, \qquad
\Delta = \frac{v-c}{c} = ( 2.53 \pm 0.78 \pm 0.29 ) \times 10^{-5} \,,
\end{align}
and the overall average is
\begin{align}
\label{e}
E_\nu =& \; 17 \, \GeV \,, \qquad
\delta t = (57.8 \pm 7.2 \pm 7.1) \, \ns \,, \qquad
\Delta = \frac{v-c}{c} = ( 2.37 \pm 0.32 \pm 0.29 ) \times 10^{-5} \,.
\end{align}
\end{subequations}
While the OPERA data rather point to a slight increase
in the ratio $\Delta = (v-c)/c$ with the neutrino energy,
than to a trend in the opposite direction, the data 
are generally consistent with a constant ratio 
$\Delta = (v-c)/c$ in the entire 
energy interval $13.8 \, \GeV < E_\nu < 40.7 \, \GeV$.

Tachyonic neutrinos fulfill the space-like dispersion relation $E_\nu^2 - \vec
p^{\,2} = -m_\nu^2$ and travel faster than light. Superluminality is conserved
under Lorentz boosts (see Ref.~\cite{BiDeSu1962} and Fig.~\ref{fig2} below). 
It has been argued that neutrinos traveling at
velocities consistent with the recent OPERA data should decay by neutral
massive analogues of Cerenkov radiation~\cite{CoGl2011}. The noncovariant
dispersion relation $E_\nu = |\vec p| \, v_\nu$ has been used in recent work on
the subject~\cite{CoGl2011} (here, $v_\nu$ denotes the 
neutrino velocity).  Freely propagating subluminal relativistic
particles as well as tachyons~\cite{Re2009,Bi2009,Bo2009} fulfill the
``opposite'' relation $ |\vec p| = E_\nu \, v_\nu$.  Both relations $E_\nu =
|\vec p| \, v_\nu$ and $|\vec p| = E_\nu\,v_\nu$ lead to a large virtuality $|
E_\nu^2 - \vec p^{\,2}|$ on the order of $(117\, \MeV)^2$ when applied to the
recently measured OPERA data [see Eqs.~\eqref{mmE1} and~\eqref{mmE2} below]. 
These observations are inconsistent with  beta decay end point
measurements~\cite{RoEtAl1991,AsEtAl1994,StDe1995,AsEtAl1996,%
WeEtAl1999,LoEtAl1999,BeEtAl2008}
which have led to values of a few $\eV^2$, 
for neutrinos in the $\keV$ energy range.
This confusing situation raises a number of questions.
Starting from the tachyonic Dirac equation, we
conclude that additional interactions, hitherto not accounted for, are required
in order to explain the OPERA data which exhibit 
a larger-than-expected  virtuality at higher energies,
or, expressed differently, an energy-dependent mass. 

At the current, early stage in the development of theoretical
models describing superluminal particles, a certain degree
of speculation cannot be avoided. For completeness, we should note 
that we neither consider models based on 
deformed special relativity~\cite{AC2000,AC2010,ACEtAl2011a,ACEtAl2011b}
nor kinematic constraints resulting from such 
models~\cite{CoGl2011,BiYiYuYu2011,CoNuSa2011,GM2011}
in any greater detail. Lorentz symmetry is conserved in our approach.

We start with a digression on the kinematic 
constraints to the observation of neutrinos along the 
OPERA baseline in Sec.~\ref{kc}.
The tachyonic Dirac equation and 
its solutions are being reviewed in Sec.~\ref{td}.
Chiral Yukawa interactions, which induce a neutrino mass
running via the renormalization group (RG), are studied in Sec.~\ref{running}.
Conclusions are reserved for Sec.~\ref{conclu}.
We always carefully distinguish between 
$|\vec p|$ and the four-vector $p$, and we use natural units with 
$\hbar = c = \epsilon_0 = 1$.

%
%
\section{Kinematic Constraints}
\label{kc}

\begin{figure}[t!]
\includegraphics[width=0.8\linewidth]{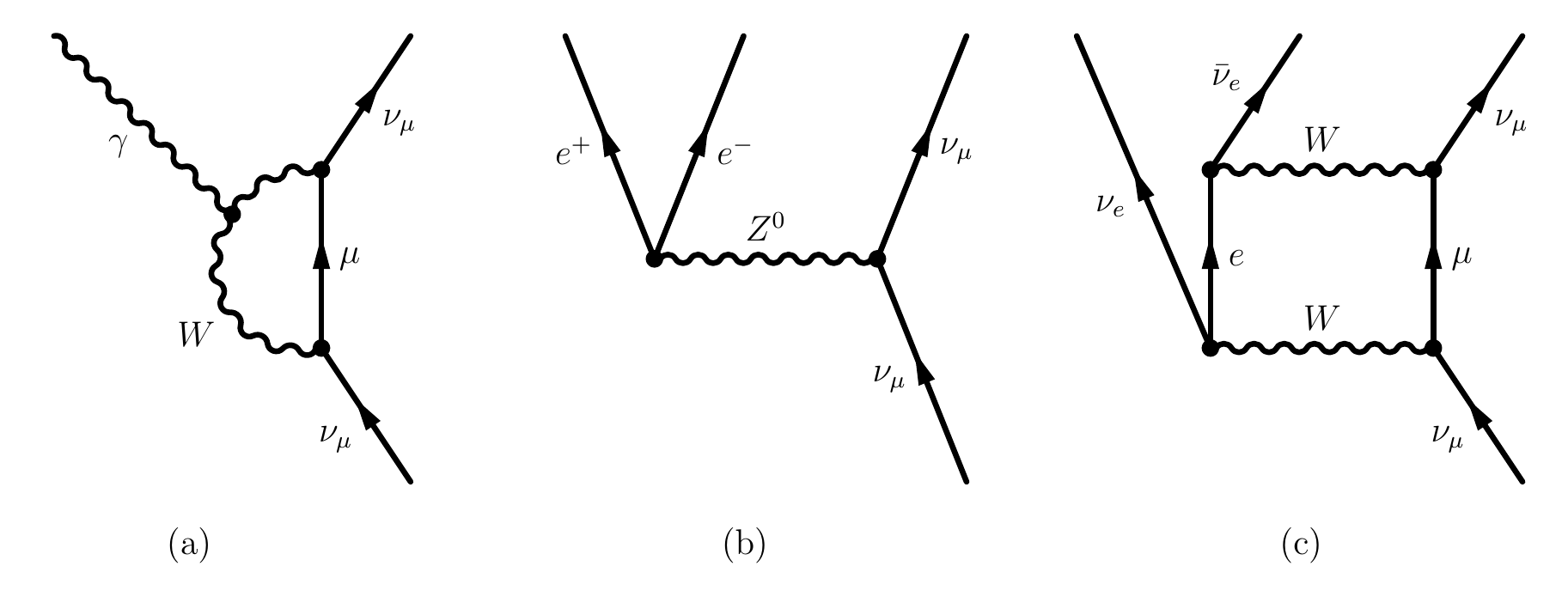}
\caption{\label{fig1}Feynman diagrams for the decay processes
of a tachyonic superluminal neutrino, as given in 
Eq.~\eqref{decays}. The tachyonic neutrino may emit 
of photon via a $W$ loop [Fig.~(a)], or an electron-positron 
pair, [Fig.~(b)], or a neutrino-antineutrino pair [Fig.~(c)].
The processes scale with the quantum electrodynamic 
(QED) coupling constant $\alpha$ 
and the weak coupling constant $G_F$ as follows,
(a)~is proportional to $\alpha \, G_F^2$,
(b)~is proportional to $\alpha \, G_F$,
and (c)~is proportional to $\alpha^2 \, G_F^2$.}
\end{figure}

The recent OPERA experiment has analyzed the propagation of muon neutrinos.
If neutrinos propagate faster than the speed of light, 
then a number of decay processes are 
kinematically allowed which are otherwise 
forbidden. These include the following decays (see Fig.~\ref{fig1}),
\begin{subequations}
\label{decays}
\begin{align}
\label{gamma}
\nu_\mu \to & \; \nu_\mu + \gamma \,, \\[0.77ex]
\label{ee}
\nu_\mu \to & \; \nu_\mu + e^+ + e^- \,, \\[0.77ex]
\label{nunu}
\nu_\mu \to & \; \nu_\mu + \nu_e + \bar\nu_e \,.
\end{align}
\end{subequations}
In Ref.~\cite{CoGl2011}, these decay processes are analyzed
under the assumption of the Lorentz-violating dispersion relation
\begin{equation}
\label{displor}
\frac{\dd E_\nu}{\dd  | \vec p_\nu | } = \mathrm{const.} \,, 
\quad
E_\nu =  | \vec p_\nu |  \, v_\nu \,,
\quad
v_\nu \approx 1 + \Delta \,,
\end{equation}
where $\Delta = 2.37 \times 10^{-5}$ corresponds to the 
value given in Ref.~\cite{OPERA2011v2}.
Processes (a)~and~(c) are parametrically suppressed with respect 
to process~(b), and therefore process~(b) is deemed to constitute
the dominant decay channel.

\begin{figure}[t!]
\includegraphics[width=0.5\linewidth]{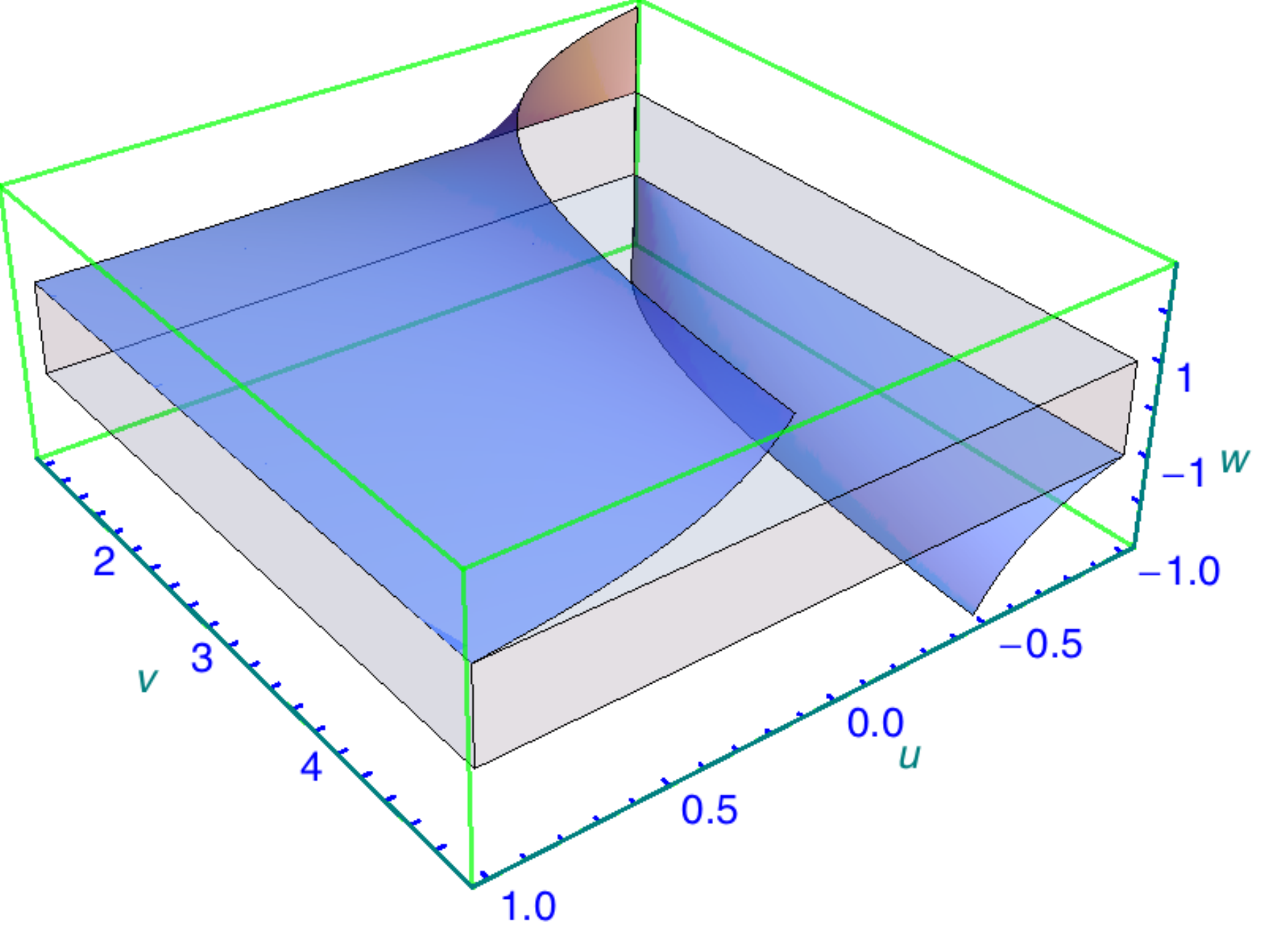}
\caption{\label{fig2}(Color online.) Illustration of the Einstein velocity
addition theorem $w = (u+v)/(1 + u \, v)$, in
the superluminal domain with $u \in (-1,1)$ and $v \in (1,3)$.
For superluminal $v$, the range $w \in [-1,1]$ of values is excluded,
as shown by the rectangular box.}
\end{figure}

One may observe that the dispersion relation 
$E_\nu = |\vec p_\nu| \, v_\nu$ is at variance with both the 
subluminal (also called tardyonic, see Ref.~\cite{Fe1967}) 
dispersion relation for freely 
propagating massive neutrinos,
\begin{equation}
\label{dispsub}
E_\nu = \frac{m_\nu}{\sqrt{1 - v_\nu^2}} \,,
\qquad
|\vec p_\nu| = \frac{m \, v_\nu}{\sqrt{1 - v_\nu^2}} = E_\nu \, v_\nu\,,
\qquad
v_\nu <  1 \,,
\end{equation}
as well as with the dispersion relation for superluminal (tachyonic)
particles~\cite{ArSu1968,DhSu1968, SuSh1986,Re2009,Bi2009,Bo2009}, 
which reads
\begin{equation}
\label{dispsup}
E_\nu = \frac{m_\nu}{\sqrt{v_\nu^2 - 1}} \,,
\qquad
|\vec p_\nu|  = \frac{m_\nu \, v_\nu}{\sqrt{v_\nu^2 - 1}} = E_\nu \, v_\nu\,,
\qquad
v_\nu > 1 \,.
\end{equation}
In both cases~\eqref{dispsub} and~\eqref{dispsup}, 
one obtains $|\vec p_\nu| = E_\nu \, v_\nu$, not the 
opposite relation $E_\nu = |\vec p_\nu| \, v_\nu$ 
used in Ref.~\cite{CoGl2011}.
Under Lorentz transformations, superluminality of tachyonic 
particles is conserved
(see Fig.~\ref{fig2}).
In two recent papers~\cite{MoRa2011,LiLiMeWaZh2011},
it has been observed that the conclusions of~\cite{CoGl2011}
would change if the dispersion relation were different.
Here, we are concerned with a more general question:
Namely, to investigate how the kinematic 
constraints change when we assume a tachyonic dispersion
relation for the neutrino, and whether the 
process~\eqref{decays} is still kinematically allowed
when $E_\nu^2 - \vec p_\nu^{\,2} < 0$.

For the process~\eqref{gamma},
an easy calculation based on the 
energy and momentum conservation conditions reveals that 
\begin{equation}
\label{kgamma}
E_\nu = E'_\nu + E_\gamma \,,
\qquad
\vec p_\nu = \vec p'_\nu + \vec k_\gamma \,,
\qquad
E_\nu = \sqrt{\vec p^{\,2}_\nu - m_\nu^2} \,,
\qquad
E'_\nu = \sqrt{\vec p'^{\,2}_\nu - m_\nu^2} \,,
\qquad
E_\gamma = | \vec k_\gamma | \,.
\end{equation}
Squaring the energy conservation condition, one obtains
\begin{subequations}
\begin{align}
\label{kgamma2}
E^2_\nu =& \;
\left( \vec p'_\nu + \vec k_\gamma \right)^{\,2} - m_\nu^2 = 
\vec p'^{\,2}_\nu + \vec k_\gamma^{\,2} 
- m_\nu^2 + 2 \, \vec p'_\nu \cdot \vec k_\gamma \,,
\\[2ex]
E^2_\nu =& \; (E'_\nu + E_\gamma)^2 =
\vec p'^{\,2}_\nu + \vec k_\gamma^{\,2} - m_\nu^2 + 
2 \, | \vec k_\gamma | \, \sqrt{ \vec p'^{\,2}_\nu - m_\nu^2}  \,,
\\[2ex]
\label{sevenc}
\vec p'_\nu \cdot \vec k_\gamma =& \;
| \vec k_\gamma | \, \sqrt{ \vec p'^{\,2}_\nu - m_\nu^2} \,.
\end{align}
\end{subequations}
We conclude that under the assumption of the 
Lorentz-covariant, tachyonic dispersion relation~\eqref{dispsup},
weak-interaction Cerenkov radiation is allowed.
In view of Eq.~\eqref{sevenc}, the photon
is radiated off at a Cerenkov angle
\begin{equation}
\label{thetagamma}
\cos \theta_\gamma = 
\frac{ \vec p'_\nu \cdot \vec k_\gamma }%
{| \vec k_\gamma | \, |\vec p'_\nu |} =
\frac{\sqrt{\vec p'^2_\nu - m_\nu^2}}{|\vec p'_\nu|} = 
\frac{E'_\nu}{|\vec p'_\nu|} = 
\frac{1}{v'_\nu} < 1 \,,
\end{equation}
under the assumption of a tachyonic neutrino with dispersion~\eqref{kgamma2}.
One may add that the kinematic consideration is somewhat 
analogous to that for the emission of ordinary Cerenkov radiation.
The important observation is that under the tachyonic 
dispersion relation~\eqref{dispsup}, the emission of a photon 
by the neutrino is always allowed, i.e., there is no threshold
energy for the neutrino and there is no threshold for the 
tachyonic mass $-m_\nu^2$. Once the particle becomes
tachyonic, weak Cerenkov radiation is kinematically allowed,
but the Cerenkov cone narrows as $-m_\nu^2 \to 0$.
For a particle fulfilling the noncovariant 
dispersion relation $E_\nu = |\vec p'_\nu| \, v_\nu$, with 
$v_\nu > 1$, the modified Cerenkov angle $\cos \theta'_\gamma$ is
easily computed as 
\begin{equation}
\cos \theta'_\gamma = 
\frac{1}{v'_\nu} + \frac{(v'^2_\nu-1) |\vec k_\gamma|}{2v'_\nu \,E'_\nu}
\approx \frac{1}{v'_\nu} < 1  \,,
\end{equation}
assuming a neutrino with the dispersion $E'_\nu = p'_\nu \, v'_\nu$ and 
$v'_\nu > 1$. This is very well approximated by $\cos \theta'_\gamma \approx 1/v'_\nu$
for $v'_\nu \approx 1$.

As a second step, let us consider a process in which 
a tachyonic neutrino fulfilling Eq.~\eqref{dispsup} 
emits a massive neutral vector meson of mass $m_0$.
This is not depicted in Fig.~\eqref{fig1} but still instructive.
The kinematic conditions change, 
\begin{equation}
\label{kmeson}
E_\nu = E'_\nu + E_0 \,,
\qquad
\vec p_\nu = \vec p'_\nu + \vec k_0 \,,
\qquad
E_\nu = \sqrt{\vec p^{\,2}_\nu - m_\nu^2} \,,
\qquad
E'_\nu = \sqrt{\vec p'^{\,2}_\nu - m_\nu^2} \,,
\qquad
E_0 = \sqrt{\vec k^{\,2}_0 + m_0^2} \,.
\end{equation}
The Cerenkov angle then becomes
\begin{equation}
\label{theta0}
\cos \theta_0 = \frac{ m_0^2 + 
2 \, \sqrt{ \vec k_0^2 + m_0^2 } \, \sqrt{ \vec p'^2_\nu - m_\nu^2}}%
{2 \, | \vec k_0 | \, |\vec p'_\nu|} 
\approx
\frac{\sqrt{ \vec k_0^2 + m_0^2 } \, \sqrt{ \vec p'^2_\nu - m_\nu^2}}%
{| \vec k_0 | \, |\vec p'_\nu|} \,,
\end{equation}
where the last expression is valid in the high-energy limit,
i.e, for $| \vec k_0 | \gg m_0$, and $|\vec p'_\nu| \gg m_\nu$.
If the vector meson carries away the bulk of the energy, 
i.e. $E_0 = x \, E_\nu$ and $E'_\nu = (1 - x) \, E_\nu$,
with $x \lesssim 1$,
then for highly energetic incoming superluminal neutrinos, 
one can always find a narrow cone near $\theta_0 \approx 0$ 
in which vector meson emission is possible. 
Again, for highly energetic tachyonic superluminal 
neutrinos, we conclude that there is no kinematic constraint
on the size of the tachyonic mass term $-m_\nu^2$ 
which would restrict massive vector meson emission.
Once the particle becomes
tachyonic and the energy of the tachyonic particle
is large enough, 
massive vector emission becomes kinematically allowed in a 
narrow angular region.
By contrast, if we replace in Eq.~\eqref{theta0}
$-m_\nu^2 \to +m_\nu^2$,
we would have $\cos \theta_0 > 1$, forbidding vector meson emission.
Also, the Cerenkov angle $\theta_0$ vanishes in the limit $m_\nu \to 0$.
Using more extensive calculations, we have checked
that the same statement applies to the light fermion pair 
emission given in Eq.~\eqref{ee} and
depicted in Fig.~\ref{fig1}(b). Cerenkov-type pair emission 
becomes kinematically possible for highly 
energetic neutrinos, in a narrow angular region.

In the application of the tachyonic dispersion relation~\eqref{dispsup}
to the OPERA data, we face a dilemma which also plagues the 
application of the Lorentz-noncovariant dispersion relation~\eqref{displor}.
Namely, we have for the OPERA data according to Eq.~\eqref{e},
\begin{align}
\label{mmE1}
-m_\nu^2 = E_\nu^2 - \vec p^{\,2}_\nu =
E_\nu^2 \, \left[ 1 - (1 + \Delta)^2 \right] = 
- \left( 117 \, \MeV \right)^2
\qquad
\mbox{[dispersion relation~\eqref{dispsup}]}
\end{align}
which is at least six orders of magnitude larger than the 
neutrino masses at 
low energy~\cite{RoEtAl1991,AsEtAl1994,AsEtAl1996,%
StDe1995,WeEtAl1999,LoEtAl1999,BeEtAl2008}.
Likewise, assuming the dispersion relation~\eqref{displor}
implies that
\begin{align}
\label{mmE2}
E_\nu^2 - \vec p^{\,2}_\nu =
\vec p^{\,2}_\nu \, \left[ (1 + \Delta)^2 - 1 \right] \approx
E^2_\nu \, \left[ (1 + \Delta)^2 - 1 \right] =
\left( 117 \, \MeV \right)^2
\quad
\mbox{[dispersion relation~\eqref{displor}]} \,.
\end{align}
If we define the expression $| E_\nu^2 - \vec p^{\,2}_\nu|$
as the ``virtuality'' of the neutrino which measures
the deviation of the neutrino propagation velocity from the speed of light,
then we can say that at the high OPERA energies, the neutrino velocity 
was not expected to deviate so much from the speed of light,
neither in the superluminal nor in the subluminal direction.
For example, if OPERA had hypothetically found a result of
\begin{equation}
\label{hypothetical}
\widetilde\Delta = -2.37 \times 10^{-5} 
\qquad
\mbox{[opposite sign as compared to Eq.~\eqref{e}]}\,,
\end{equation}
then this would have been equally surprising.
In the latter case, one would probably have concluded 
immediately that the neutrino must be subject to 
a hitherto unknown interaction at high energy, 
modifying its effective (running) mass.
We here advocate the viewpoint that the same 
conclusion should be drawn from the OPERA data: namely,
the neutrino is genuinely tachyonic and subject to an 
unknown interaction at high energy which modifies its mass 
and its decay channels. Otherwise, it seems that the 
high-energy OPERA data~\cite{OPERA2011v2} (in the 
$\GeV$ range) cannot be reconciled with the 
low-energy experimental
results (in the $\keV$ range) of Refs.~\cite{RoEtAl1991,AsEtAl1994,AsEtAl1996,StDe1995,%
WeEtAl1999,LoEtAl1999,BeEtAl2008}.
Of course, this statement holds provided the OPERA data 
are not subject to a hitherto undiscovered systematic error.

The data bins given in Eq.~\eqref{eee}
are consistent with an energy-independent propagation 
velocity. While the quantity $\Delta$ need not be 
energy independent over large energy intervals, 
it appears to be so in in the energy interval
$13.8 \, \GeV < E_\nu < 40.7 \, \GeV$.
Therefore, in this energy interval observed by OPERA~\cite{OPERA2011v2},
the dispersion relation is assumed to be close to a linear relationship 
\begin{equation}
\label{mmE}
m_\nu = m(E_\nu) \approx \eta \, E_\nu \,,
\qquad
13.8 \, \GeV < E_\nu < 40.7 \, \GeV \,,
\qquad
\eta = \sqrt{(1 + \Delta)^2 - 1} = 6.88\times 10^{-3} \approx \frac{1}{145} \,.
\end{equation}
The unknown interaction leading to the energy-dependent mass 
must now be investigated. When calculating
decay rates, the existence of the additional
interaction implies that one should use eigenstates of the 
neutrino in the additional, hitherto unknown 
interaction potential (i.e., taking into account the running mass) 
rather than freely propagating tachyonic states, with an effective 
energy-dependent tachyonic mass $m_\nu = m_\nu(E_\nu) \propto E_\nu$.

%
%
\section{Tachyonic Dirac Equation}
\label{td}

Given the obvious inconsistency of the OPERA data~\cite{OPERA2011v2}
with low-energy neutrino data~\cite{RoEtAl1991,AsEtAl1994,AsEtAl1996,StDe1995,%
WeEtAl1999,LoEtAl1999,BeEtAl2008},
as manifest in the energy-dependent effective mass~\eqref{mmE},
one may ask why an equation that describes a genuinely 
tachyonic neutrino with an energy-independent, fixed
tachyonic mass $m_\nu$
should be considered at all in the following.
The reason is that if the neutrino is genuinely tachyonic, 
then one has to start from an equation which describes 
a genuinely tachyonic particle, with the possibility to 
describe additional perturbative 
interactions that modify the high-energy behaviour.
Expressed differently, we would expect the tachyonic Dirac 
equation given below to describe the low-energy behaviour
of neutrinos~\cite{RoEtAl1991,AsEtAl1994,AsEtAl1996,StDe1995,%
WeEtAl1999,LoEtAl1999,BeEtAl2008},
while the large deviation from the light cone seen at high 
energies~\cite{AdEtAl2007,OPERA2011v2} should be ascribed to 
additional interactions.
We briefly recall here that 
the Lorentz-covariant tachyonic Dirac equation reads
\begin{equation}
\label{Dirac5}
\left( \ii \gamma^\mu \partial_\mu - 
\gamma^5 \,m_\nu\right) \psi(x) = 0  \,,
\qquad
\gamma^0 =\left( \begin{array}{cc} \mathbbm{1}_{2\times 2} & 0 \\ 
0 & -\mathbbm{1}_{2\times 2} \\
\end{array} \right) \,,
\qquad
\vec\gamma = \left( \begin{array}{cc} 0 & \vec\sigma \\ -\vec\sigma & 0  \\
\end{array} \right) \,,
\qquad
\gamma^5 = \left( \begin{array}{cc} 0 & \mathbbm{1}_{2\times 2} \\ 
\mathbbm{1}_{2\times 2} & 0  \\
\end{array} \right) \,.
\end{equation}
Here, $x = (t, \vec x)$, and we use the Dirac matrices
in the Dirac representation~\cite{JeWu2011jpa}.
The partial derivatives are $\partial_\mu = \partial/\partial x^\mu$, 
while $\gamma^5 = \ii \, \gamma^0 \, \gamma^1 \, \gamma^2 \, \gamma^3$
is the fifth current matrix. The tachyonic Dirac equation has been briefly discussed
in Ref.~\cite{ChHaKo1985,Ch2000,Ch2002}.
It has recently been verified that 
this equation is $\calC \cal P$, as well as $\calT$ invariant~\cite{JeWu2011jpa}.
These symmetry properties apply to neutrinos.
The positive-energy plane-wave solutions~\cite{JeWu2011jpa} 
of the tachyonic Dirac equation have the properties 
\begin{equation}
\label{solutions}
\Psi(x) = \frac{1}{\sqrt{V}} U_\pm(\vec k_\nu) \, \ee^{-\ii k_\nu \cdot x} \,,
\quad
k_\nu = (E_\nu, \vec k_\nu) \,,
\quad
E_\nu = \sqrt{\vec k_\nu^{2} - m_\nu^2} \,, \quad | \vec k_\nu | \geq m_\nu \,.
\end{equation}
The negative-energy solutions~\cite{JeWu2011jpa} are given by
\begin{equation}
\Phi(x) = \frac{1}{\sqrt{V}} V_\pm(\vec k_\nu) \, \ee^{\ii k_\nu \cdot x} \,,
\quad
k_\nu = (E_\nu, \vec k_\nu) \,,
\quad
E_\nu = \sqrt{\vec k_\nu^2 - m_\nu^2} \,,
\quad |\vec k_\nu| \geq m_\nu \,,
\end{equation}
where $V$ is the normalization volume.
These states are normalized, with
$U^\plus_+(\vec k_\nu) \, U_+(\vec k_\nu) =
U^\plus_-(\vec k_\nu) \, U_-(\vec k_\nu) =
V^\plus_+(\vec k_\nu) \, V_+(\vec k_\nu) =
V^\plus_-(\vec k_\nu) \, V_-(\vec k_\nu) = 1$.
The spinors entering these expressions read as
\begin{equation}
\label{UU}
U_+(\vec k_\nu) = 
\left( \begin{array}{c}
\dfrac{m_\nu-E_\nu+|\vec k_\nu|}{\sqrt{2} \, 
\sqrt{(E_\nu - |\vec k_\nu|)^2 + m_\nu^2}} \; a_+(\vec k_\nu) \\[0.77ex]
\dfrac{m_\nu+E_\nu-|\vec k_\nu|}{\sqrt{2} \, 
\sqrt{(E_\nu - |\vec k_\nu|)^2 + m_\nu^2}} \; a_+(\vec k_\nu) \\
\end{array} \right) \,,
\qquad
U_-(\vec k_\nu) = 
\left( \begin{array}{c}
\dfrac{m_\nu+E_\nu-|\vec k_\nu|}{\sqrt{2} \, 
\sqrt{(E_\nu - |\vec k_\nu|)^2 + m_\nu^2}} \; 
a_-(\vec k_\nu) \\[0.77ex]
\dfrac{-m_\nu+E_\nu-|\vec k_\nu|}%
{\sqrt{2} \, \sqrt{(E_\nu - |\vec k_\nu|)^2 + m_\nu^2}} \; 
a_-(\vec k_\nu) \\
\end{array} \right) \,,
\end{equation}
where the helicity spinors $a_\pm(\vec k_\nu)$ are given below
in Eq.~\eqref{aplusminus}.
If we are interested in the massless limit, then we 
should first take into account the fact that massless
particles propagate at velocities very close to the light cone.
For $v = 1 + \Delta$, we have $E - |\vec k_\nu| \approx -m \, \Delta/2 
\ll m$.
Therefore, letting $\Delta \to 0$, 
the dominant term for the massless limit actually 
is the mass $m \gg E - |\vec k_\nu|$. This implies, e.g., that 
$U_+(\vec k_\nu) \to \frac{1}{\sqrt{2}} 
\left( \begin{array}{c} a_+(\vec k_\nu) \\
a_+(\vec k_\nu) \\ \end{array} \right)$ for the massless case.
The negative-energy eigenstates are given by
\begin{equation}
\label{VV}
V_+(\vec k_\nu) = 
\left( \begin{array}{c}
\dfrac{-m_\nu-E_\nu+|\vec k_\nu|}%
{\sqrt{2} \, \sqrt{(E_\nu - |\vec k_\nu|)^2 + m_\nu^2}} \; 
a_+(\vec k_\nu) \\[0.77ex]
\dfrac{-m_\nu+E_\nu-|\vec k_\nu|}%
{\sqrt{2} \, \sqrt{(E_\nu - |\vec k_\nu|)^2 + m_\nu^2}} \; 
a_+(\vec k_\nu) \\
\end{array} \right) \,,
\qquad
V_-(\vec k_\nu) =
\left( \begin{array}{c}
\dfrac{-m_\nu+E_\nu-|\vec k_\nu|}%
{\sqrt{2} \, \sqrt{(E_\nu - |\vec k_\nu|)^2 + m_\nu^2}} \; 
a_-(\vec k_\nu) \\[0.77ex]
\dfrac{m_\nu+E_\nu-|\vec k_\nu|}%
{\sqrt{2} \, \sqrt{(E_\nu - |\vec k_\nu|)^2 + m_\nu^2}} \; a_-(\vec k_\nu) \\
\end{array} \right) \,.
\end{equation}
The helicity spinors entering these expressions are
given in terms of the polar and azimuthal angles 
$\theta$ and $\varphi$ of the three-vector $\vec k_\nu$,
\begin{equation}
\label{aplusminus}
a_+(\vec k_\nu) = \left( \begin{array}{c}
\cos\left(\frac{\theta}{2}\right) \\[1ex]
\sin\left(\frac{\theta}{2}\right) \, \ee^{\ii \, \varphi} \\
\end{array} \right) \,,
\qquad
a_-(\vec k_\nu) = \left( \begin{array}{c}
-\sin\left(\frac{\theta}{2}\right) \, \ee^{-\ii \, \varphi} \\[1ex]
\cos\left(\frac{\theta}{2}\right) \\
\end{array} \right) \,,
\end{equation}
and fulfill
\begin{equation}
\frac{\vec \sigma \cdot \vec k_\nu}{|\vec k_\nu|} \, 
a_+(\vec k_\nu) = a_+(\vec k_\nu) \,,
\qquad
\frac{\vec \sigma \cdot \vec k_\nu}{|\vec k_\nu|} \, 
a_-(\vec k_\nu) = -a_+(\vec k_\nu) \,.
\end{equation}
For plane waves, $E_\nu = \sqrt{\vec k_\nu^{\,2} - m_\nu^2}$ and 
$\vec p_\nu = \vec k_\nu$ fulfill the tachyonic dispersion relation~\eqref{dispsup},
which we recall for convenience,
\begin{equation}
\label{recall}
E_\nu = \frac{m_\nu}{\sqrt{v_\nu^2 - 1}} \,,
\qquad
|\vec k_\nu| = \frac{m \, v_\nu}{\sqrt{v_\nu^2 - 1}} = E_\nu \, v_\nu\,,
\qquad
v_\nu > 1 \,,
\end{equation}
so that $\sqrt{\vec k_\nu^{\,2} - m_\nu^2}$ never becomes 
imaginary. For $\vec k_\nu^{\,2} < m_\nu^2$, we have resonance and antiresonance
energies. We start with the resonances, whose energies have a negative 
imaginary part,
\begin{subequations}
\label{RR}
\begin{align}
R_+(\vec k_\nu) = & \;
\left( \begin{array}{c}
\dfrac{m_\nu+\tfrac{\ii}{2} \Gamma_\nu +|\vec k_\nu|}{\sqrt{2} \, 
\sqrt{\vec k_\nu^{\,2} + m_\nu^2 + \tfrac14 \, \Gamma_\nu^2}} \; 
a_+(\vec k_\nu) \\[0.77ex]
\dfrac{m_\nu-\tfrac{\ii}{2} \Gamma_\nu -|\vec k_\nu|}{\sqrt{2} \, 
\sqrt{\vec k_\nu^{\,2} + m_\nu^2 + \tfrac14 \, \Gamma_\nu^2}} \; 
a_+(\vec k_\nu) \\
\end{array} \right) \,,
\qquad
R_-(\vec k_\nu) = 
\left( \begin{array}{c}
\dfrac{m_\nu-\tfrac{\ii}{2} \Gamma_\nu -|\vec k_\nu|}{\sqrt{2} \, 
\sqrt{\vec k_\nu^{\,2} + m_\nu^2 + \tfrac14 \, \Gamma_\nu^2}} \; 
a_-(\vec k_\nu) \\[0.77ex]
\dfrac{-m_\nu-\tfrac{\ii}{2} \Gamma_\nu-|\vec k_\nu|}{\sqrt{2} \, 
\sqrt{\vec k_\nu^{\,2} + m_\nu^2 + \tfrac14 \, \Gamma_\nu^2}} \; 
a_-(\vec k_\nu) \\
\end{array} \right) \,,
\\[0.77ex]
E_\nu =& \; -\tfrac{\ii}{2}\, \Gamma_\nu = -\ii\, 
\sqrt{m_\nu^2 - \vec k_\nu^{2}} \,,
\qquad
\vec k_\nu^{\,2} < m_\nu^2 \,.
\end{align}
\end{subequations}
The antiresonance energies have a positive imaginary part,
\begin{subequations}
\label{SS}
\begin{align}
S_+(\vec k_\nu) = & \;
\left( \begin{array}{c}
\dfrac{-m_\nu-\tfrac{\ii}{2} \Gamma_\nu+|\vec k_\nu|}{\sqrt{2} \, 
\sqrt{\vec k_\nu^{\,2} + m_\nu^2 + \tfrac14 \, \Gamma_\nu^2}} \; 
a_+(\vec k_\nu) \\[0.77ex]
\dfrac{-m_\nu+\tfrac{\ii}{2} \Gamma_\nu-|\vec k_\nu|}{\sqrt{2} \, 
\sqrt{\vec k_\nu^{\,2} + m_\nu^2 + \tfrac14 \, \Gamma_\nu^2}} \; 
a_+(\vec k_\nu) \\
\end{array} \right) \,,
\qquad
S_-(\vec k_\nu) = 
\left( \begin{array}{c}
\dfrac{-m_\nu+\tfrac{\ii}{2}\Gamma_\nu -|\vec k_\nu|}{\sqrt{2} \, 
\sqrt{\vec k_\nu^{\,2} + m_\nu^2 + \tfrac14 \, \Gamma_\nu^2}} \; 
a_-(\vec k_\nu) \\[0.77ex]
\dfrac{m_\nu+\tfrac{\ii}{2}\Gamma_\nu -|\vec k_\nu|}{\sqrt{2} \, 
\sqrt{\vec k_\nu^{\,2} + m_\nu^2 + \tfrac14 \, \Gamma_\nu^2}} \;
a_-(\vec k_\nu) \\
\end{array} \right) \,,
\\[0.77ex]
E_\nu =& \; \tfrac{\ii}{2}\, \Gamma_\nu = 
\ii \, \sqrt{m_\nu^2 - \vec k_\nu^{2}} \,,
\qquad \vec k_\nu^{\,2} < m_\nu^2 \,.
\end{align}
\end{subequations}
These states are also normalized, with $R^\plus_+(\vec k_\nu) \, R_+(\vec k_\nu) =
R^\plus_-(\vec k_\nu) \, R_-(\vec k_\nu) = S^\plus_+(\vec k_\nu) \, S_+(\vec k_\nu) =
S^\plus_-(\vec k_\nu) \, S_-(\vec k_\nu) = 1$.
The term ``resonances'' is used in the physics literature 
in two contexts: (i)~in order to designate the 
complex energy eigenvalue of a Hamiltonian, 
and (ii)~in order to designate the peak in a cross section
or a quantum state which can decay into a final state
with a different particle content. In the current case,
the interpretation~(i) is relevant. The resonances have complex
resonance energies; the waves are evanescent 
(exponentially damped) just like the diffracted wave under total reflection, 
or a wave in a waveguide below the minimum frequency 
for the TE$_{1,0}$ mode necessary for propagation,
and the resonance energies are complex just as in the 
case of a resonance energy of the Stark effect~\cite{Je2001pra}.
Resonances are damped for propagation forward in time,
antiresonances for propagation backward in time, 
in accordance with the Feynman prescription.
The wavelength of the resonance states is too long 
to be supported in a genuinely superluminal wave packet of 
tachyonic mass $m_\nu^2$.

The noncovariant, Hamiltonian form of Eq.~\eqref{Dirac5}
reads as
\begin{equation}
H_5 \psi(\vec x) = 
\left( \vec \alpha \cdot \vec p + \beta \, \gamma^5 \, m_\nu \right) \, \psi(\vec x) = 
E_\nu\, \psi(\vec x) \,,
\end{equation}
where $\beta = \gamma^0$, and $\vec \alpha = \gamma^0 \, \vec\gamma$.
The Hamiltonian $H_5$ has the 
pseudo-Hermitian~\cite{Pa1943,BeBo1998,BeDu1999,BeBoMe1999,%
BeBrJo2002,Mo2002i,Mo2002ii,Mo2002iii,Mo2003npb} property 
\begin{equation}
H 
= \calP \, H_5^\plus(\vec x) \, \calP^{-1} 
= P \, H_5^\plus(-\vec x) \, P^{-1} \,,
\end{equation}
where $\calP$ is the full parity transformation and
$P$ is the parity matrix $P = \gamma^0$.  
The eigenvalues of a
pseudo-Hermitian operator come in complex-conjugate pairs and are real if the
tachyonic dispersion relations~\eqref{dispsup} are fulfilled.
This can be seen as follows. Because the spectrum of a Hermitian
adjoint operator consists of the complex conjugate eigenvalues, 
we have an eigenvector $\phi(\vec x)$ with eigenvalue 
$E^*$ provided there exists an eigenvector $\psi(\vec x)$ with 
eigenvalue $E$,
\begin{equation}
H_5(\vec x) \, \psi(\vec x) = E \, \psi(\vec x) \,,
\qquad
H_5^\plus(\vec x) \, \phi(\vec x) = E^* \, \phi(\vec x) \,.
\end{equation}
Then, the transformation $\vec x \to - \vec x$ and
the introduction of the $P =\gamma^0$ parity matrix leads to 
\begin{equation}
H_5^+(-\vec x) \, \phi(-\vec x) = E^* \, \phi(-\vec x) \,,
\qquad
P H_5^+(-\vec x) P^{-1} \, \left( P \phi(-\vec x) \right) = E^* \, P \phi(-\vec x) \,.
\end{equation}
By assumption, $P H_5^+(-\vec x) P^{-1} = H_5(\vec x)$ and thus
\begin{equation}
H_5(\vec x) \, P \phi(-\vec x) = E^* \, P \phi(-\vec x) \,,
\qquad
H_5(\vec x) \, \widetilde\psi(\vec x) = E^* \widetilde\psi(\vec x) \,,
\qquad
\widetilde\psi(\vec x) = P \phi(-\vec x) \,.
\end{equation}
This implies that $\widetilde\psi(\vec x) = P \psi(-\vec x)$ is an eigenvector
with eigenvalue $E^*$. The
eigenvalues of $H_5$ thus come in complex-conjugate pairs,
and furthermore, they are real for plane waves 
fulfilling the dispersion relation~\eqref{dispsup}.

The covariant Green function corresponding to the Hamiltonian 
$H_5$ thus reads as 
\begin{equation}
S_T(p) = \gamma^0 \, \frac{1}{E - H_5} = 
\frac{\cancel{p} + \gamma^5 \, m_\nu }{p^2 + m_\nu^2} \,.
\end{equation}
The tachyonic poles at $E_\nu^2 - \vec p^{\,2} = -m_\nu^2$
have to be encircled in a way consistent with the 
boundary conditions imposed on the Green function.
Eigenvalues with 
$E_\nu^2  = \vec p^{\,2} - m_\nu^2 < 0$
represent evanescent waves.
If one encircles the poles of the Green function
according to the Feynman prescription,
\begin{equation}
\label{ST}
S_T(p) = \frac{1}{\cancel{p} - \gamma^5 \, (m_\nu + \ii\,\epsilon)} =
\frac{\cancel{p} - \gamma^5 \, m_\nu}{p^2 + m^2_\nu + \ii \, \epsilon} \,,
\end{equation}
then the 
energy-momentum dispersion relation is infinitesimally
displaced to read
$E_\nu = \pm \sqrt{\vec p^{\,2} - m_\nu^2 - \ii \, \epsilon} $.
This is consistent with the evanescent wave picture 
because positive-energy solutions have the form
$E_\nu = \epsilon -\ii \, \sqrt{|\vec p^{\,2} - m_\nu^2|}$
and are thus exponentially damped for the propagation into the future,
whereas negative-energy solutions have the form
$E_\nu = -\epsilon +\ii \, \sqrt{|\vec p^{\,2} - m_\nu^2|}$
and are thus exponentially damped for the propagation into the past.
In general, the Feynman prescription assigns an infinitesimal 
negative imaginary part to energies whose real part is 
positive, and vice versa. 

Thus, while the time propagation of strictly tachyonic 
wave packets (superpositions of the tachyonic plane-wave solutions)
is fully unitary (they have real eigenvalues),
a slight violation of unitarity cannot be avoided 
if one allows eigenstates with $E_\nu^2 = \vec p^{\,2} - m_\nu^2 < 0$.
The complex resonance energies (the real part is only infinitesimal)
\begin{equation}
E_\nu = \epsilon -\ii \sqrt{| \vec p^{\,2} - m_\nu^2 |} \,,
\qquad
E_\nu = -\epsilon + \ii \sqrt{| \vec p^{\,2} - m_\nu^2 |} \,,
\qquad
\vec p^{\,2} < m_\nu^2 \,,
\end{equation}
describe the suppression of subluminal components of a
superluminal wave packet under time evolution.
One has to allow these solutions in the propagator~\eqref{ST}
if one would like to carry out the Fourier transformation 
consistently, i.e., over the entire range $p^\mu \in \mathbbm{R}^4$,
or describe the time evolution of a general 
wave packet under the Green function~\eqref{ST}.
It seems that a slight violation of unitarity,
relevant to the small sector $\vec p^{\,2} < m_\nu^2$,
where $m_\nu$ initially is on the order of a few $\eV$, 
is a price for the introduction of tachyonic particles~\cite{XiJi1987}.
Note that full unitarity cannot be preserved anyway in 
a tachyonic theory if one goes beyond tree-level amplitudes,
as shown in Ref.~\cite{Bo1970}.
The time propagation of wave packets in potentials 
with manifestly complex resonance energies
has been described in Refs.~\cite{MoiseyevMcCurdy,JeSuLuZJ2008}.
The evanescence of the subluminal neutrino 
wave function components, which are excluded from the 
real neutrino plane-wave eigenstates but included in the 
propagator, is somewhat analogous 
to the photon propagator, where one includes the so-called 
scalar and longitudinal photons in the photon propagator but 
leaves them out from the real, physical states of the photon 
field, which are composed of transverse photons.

In Ref.~\cite{JeWu2011jpa}, 
the tachyonic propagator~\eqref{ST} is derived not by inversion
of the Hamiltonian, but by a quantization of the tachyonic
field operators. We briefly sketch the essential elements of the derivation.
The field operator is written as
\begin{align}
\hat\psi(x) =& \;
\int \frac{\dd^3 k_\nu}{(2\pi)^3} \, 
\frac{m_\nu}{E_\nu} \sum_{\sigma = \pm} 
\left[ b_\sigma(k_\nu) \, \calU_\sigma(\vec k_\nu) \, 
\ee^{-\ii \, k_\nu \cdot x} 
+ b_\sigma(-k_\nu) \, \calV_\sigma(\vec k_\nu) \, 
\ee^{\ii \, k_\nu \cdot x} \right] 
\nonumber\\[2ex]
=& \;
\int \frac{\dd^3 k_\nu}{(2\pi)^3} \, 
\frac{m}{E_\nu} \sum_{\sigma = \pm} 
\left[ b_\sigma(k_\nu) \, \calU_\sigma(\vec k_\nu) \, 
\ee^{-\ii \, k_\nu \cdot x} 
+ d^\plus_\sigma(k_\nu) \, \calV_\sigma(\vec k_\nu) \, 
\ee^{\ii \, k_\nu \cdot x} \right] \,,
\end{align}
where $E_\nu =  \sqrt{\vec k_\nu^2 - m_\nu^2 - \ii \, \epsilon}$
is the tachyonic energy and the four-vector $k_\nu$ equals
$k_\nu = (E_\nu, \vec k_\nu)$.
Here, the $b$ operators annihilate particle,
whereas the $d^\plus$ create antiparticles.
We here explicitly accept a Lorentz-covariant 
vacuum state, which transforms according to Ref.~\cite{Fe1967}.
The Lorentz-transformed vacuum is filled with all 
particle and antiparticle states whose energy changes
sign under a Lorentz transformation (Lorentz boost), 
as outlined in Eqs.~(5.6) and (5.7) of Ref.~\cite{Fe1967}.
The spinors $\calU$ and $\calV$ are given by
\begin{subequations}
\label{covariant}
\begin{equation}
\calU_+(\vec k_\nu) = 
\left( \begin{array}{c}
\dfrac{m_\nu-E_\nu+|\vec k_\nu|}{2 \, \sqrt{m_\nu} \, \sqrt{|\vec k_\nu|- m_\nu}} \; 
a_+(\vec k_\nu) \\[0.77ex]
\dfrac{m_\nu+E_\nu-|\vec k_\nu|}{2 \, \sqrt{m_\nu} \, \sqrt{|\vec k_\nu|- m_\nu}} \; 
a_+(\vec k_\nu) \\
\end{array} \right) \,,
\qquad
\calU_-(\vec k_\nu) =
\left( \begin{array}{c}
\dfrac{m_\nu+E_\nu-|\vec k_\nu|}{2 \, \sqrt{m_\nu} \, \sqrt{|\vec k_\nu|- m_\nu}} \; 
a_-(\vec k_\nu) \\[0.77ex]
\dfrac{-m_\nu+E_\nu-|\vec k_\nu|}{2 \, \sqrt{m_\nu} \, \sqrt{|\vec k_\nu|- m_\nu}} \; 
a_-(\vec k_\nu) \\
\end{array} \right) 
\end{equation}
for positive energy, and by 
\begin{equation}
\calV_+(\vec k_\nu) = 
\left( \begin{array}{c}
\dfrac{-m_\nu-E_\nu+|\vec k_\nu|}{2 \, \sqrt{m_\nu} \, \sqrt{|\vec k_\nu|- m_\nu}} \; 
a_+(\vec k_\nu) \\[0.77ex]
\dfrac{-m_\nu+E_\nu-|\vec k_\nu|}{2 \, \sqrt{m_\nu} \, \sqrt{|\vec k_\nu|- m_\nu}} \; 
a_+(\vec k_\nu) \\
\end{array} \right) \,,
\qquad
\calV_-(\vec k_\nu) = 
\left( \begin{array}{c}
\dfrac{-m_\nu+E_\nu-|\vec k_\nu|}{2 \, \sqrt{m_\nu} \, \sqrt{|\vec k_\nu|- m_\nu}} \; 
a_-(\vec k_\nu) \\[0.77ex]
\dfrac{m_\nu+E_\nu-|\vec k_\nu|}{2 \, \sqrt{m_\nu} \, \sqrt{|\vec k_\nu|- m_\nu}} \; 
a_-(\vec k_\nu) \\
\end{array} \right) 
\end{equation}
for negative energy (in both cases, we assume
that $|\vec k_\nu| > m$). The normalization conditions are given by
\begin{equation}
\overline \calU_\sigma(\vec k_\nu) \; \calU_\sigma(\vec k_\nu) =
\calU^\plus_\sigma(\vec k_\nu) \gamma^0 \calU_\sigma(\vec k_\nu) = \sigma \,,
\qquad
\overline \calV_\sigma(\vec k_\nu) \; \calV_\sigma(\vec k_\nu) =
\calV^\plus_\sigma(\vec k_\nu) \gamma^0 \calV_\sigma(\vec k_\nu) = -\sigma \,.
\end{equation}
\end{subequations}
Quantizing the theory according to Fermi--Dirac statistics,
\begin{subequations}
\label{quantization}
\begin{align}
\left\{ b_\sigma(k_\nu) , b_{\rho}(k_\nu') \right\} = & \;
\left\{ a^\plus_\sigma(k_\nu) , a^\plus_{\rho}(k_\nu') \right\} = 
\left\{ d_\sigma(k_\nu) , d_{\rho}(k_\nu') \right\} =
\left\{ d^\plus_\sigma(k_\nu) , d^\plus_{\rho}(k_\nu') \right\} = 0 \,, 
\\[0.77ex]
\left\{ b_\sigma(k_\nu) , b^\plus_{\rho}(k_\nu') \right\} = & \; 
(-\sigma) \, 
(2 \pi)^3 \, \frac{E}{m} \delta^3(\vec k_\nu - \vec k_\nu') \, \delta_{\sigma\rho}\,,
\qquad
\left\{ d_\sigma(k_\nu) , d^\plus_{\rho}(k_\nu') \right\} = 
(-\sigma) \, 
(2 \pi)^3 \, \frac{E}{m} \delta^3(\vec k_\nu - \vec k_\nu') \, 
\delta_{\sigma\rho}\,,
\end{align}
\end{subequations}
one can easily show that
\begin{equation}
\label{tensor}
\sum_\sigma (-\sigma) \; \calU_\sigma(\vec k_\nu) \otimes
\overline\calU_\sigma(\vec k_\nu) \,\gamma^5 =
\frac{\cancel{k}_\nu - \gamma^5 \, m}{2 m} \,,
\qquad
\qquad
\sum_\sigma (-\sigma) \; \calV_\sigma(\vec k_\nu) \otimes
\overline\calV_\sigma(\vec k_\nu) \,\gamma^5 =
\frac{\cancel{k}_\nu + \gamma^5 \, m}{2 m} \,.
\end{equation}
The field anticommutator is 
\begin{align}
& \{ \hat\psi_\xi(x), \overline{\hat\psi}_{\xi'}(y) \} = 
\left< 0 \left| \{ \hat\psi_\xi(x), 
\overline{\hat\psi}_{\xi'}(y) \} \right| 0 \right> 
\nonumber\\[2ex]
& \qquad = \int \frac{\dd^3 k_\nu}{(2 \pi)^3} 
\frac{m_\nu}{E_\nu} \,
\sum_{\sigma = \pm} \left\{
\ee^{-\ii k_\nu \cdot (x-y)} \,
\left(-\sigma\right) \, \left[ \calU_{\sigma}(\vec k_\nu) \right]_\xi \, 
\left[ \overline \calU_{\sigma}(\vec k_\nu) \right]_{\xi'} 
+
\ee^{\ii k_\nu \cdot (x-y)} \,
\left(-\sigma\right) \, \left[ \calV_{\sigma}(\vec k_\nu) \right]_\xi \, 
\left[ \overline \calV_{\sigma}(\vec k_\nu) \right]_{\xi'} \right\} \,,
\end{align}
where $\xi$ denotes the spinor index.
It follows that
\begin{equation}
\label{anticom2}
\{ \hat\psi_\xi(x), \overline{\hat\psi}_{\xi'}(y) \} \, \gamma^5 = 
\left( \ii \, \cancel{\partial} - \gamma^5 \, m_\nu \right)_{\xi \xi'} 
\ii \, \Delta(x - y) \,,
\qquad
\qquad
\Delta(x-y) =
-\ii \int \frac{\dd^3 k_\nu}{(2 \pi)^3} \, 
\frac{1}{2 E_\nu} \, \left( \ee^{-\ii k_\nu \cdot (x-y)} -
\ee^{\ii k_\nu \cdot (x-y)} \right)\,,
\end{equation}
where $\Delta(x-y)$ is introduced as in Chap.~3 of Ref.~\cite{ItZu1980}.
Furthermore, Eq.~(3.170) of~\cite{ItZu1980} finds the generalization
\begin{equation}
\left. 
\{ \hat\psi_\xi(x), \overline{\hat\psi}_{\xi'}(y) \} \, \gamma^5 
\right|_{x_0 = y_0} 
= -
\left. 
\left( \gamma^0 \right)_{\xi\xi'} \, \partial_0 \, \Delta(x - y)
\right|_{x_0 = y_0} 
= \left( \gamma^0 \right)_{\xi\xi'} \, \delta^3(\vec x - \vec y) \,.
\end{equation}
In full analogy with Eq.~(3.174) of Ref.~\cite{ItZu1980} 
and in agreement with Ref.~\cite{JeWu2012},
the tachyonic ($T$) propagator is then found as
\begin{equation}
\left< 0 \left| T \, \hat\psi_\xi(x) \, 
\overline{\hat\psi}_{\xi'}(y) \gamma^5 \right| 0 \right> = 
\ii \, S_T(x - y)_{\xi\xi'} \,,
\qquad
\qquad
S_T(x - y) = 
\int \frac{\dd^4 k_\nu}{(2 \pi)^4} \, 
\ee^{-\ii k_\nu \cdot (x-y)} \,
\frac{\cancel{k}_\nu - \gamma^5 \, m_\nu}{k_\nu^2 + m^2_\nu + \ii \, \epsilon} \,,
\end{equation}
which confirms Eq.~\eqref{ST}. Indeed, the propagator obtained 
from the quantized theory is equal to the propagator obtained 
from the inversion of the Hamiltonian in the Lorentz-covariant
formulation, as it should.
The couplings of the neutrino involve the chirality 
projector $(1 - \gamma^5)/2$, and in view 
of $\gamma^5 \, (1 \pm \gamma^5)/2 = \pm (1 \pm \gamma^5)/2$,
the introduction of the $\gamma^5$ matrix in Eq.~\eqref{ST} is
reabsorbed into the interaction Lagrangian.
The non-unitarity is small because $m_\nu^2 $ is very small.
For a tachyonic particle, the evanescence 
of non-tachyonic wave packet components is natural because its
tachyonic components remain tachyonic upon
Lorentz transformation (see Fig.~\ref{fig2}).
The tachyonic Dirac equation provides for a convenient 
framework to describe freely propagating, superluminal, 
electromagnetically neutral, particles.

%
%
\section{Neutrino Mass Running}
\label{running}

In Secs.~\ref{kc} and~\ref{td}, we have seen that a running neutrino
mass (with the energy) is able to conceivably suppress
Cerenkov-type decay processes, and the quantization of the 
tachyonic Dirac equation has been discussed as a convenient description for 
tachyonic spin-$1/2$ particles;
it naturally implies the suppression of the right-handed neutrino.
If current experimental data~\cite{OPERA2011v2}
are confirmed, then we now have to explain why the
effective mass of the neutrino, which needs to be inserted into the 
tachyonic Dirac equation, changes from a few eV in the keV neutrino energy  range, 
to a mass on the order of MeV in the GeV energy range.
We note that neutrino mass running is usually assumed 
to initiate on the energy scales of Grand Unification (see-saw mechanism).
However, the experimental data~\cite{DaEtAl1987,AdEtAl2007,OPERA2011v2,%
RoEtAl1991,AsEtAl1994,AsEtAl1996,StDe1995}
all point to a neutrino mass running which sets in at much 
lower energy scales. We assume that the mass term is genuinely tachyonic.

The scenario that we would like to propose is as follows:
We conjecture that the neutrino mass running is due to 
an interaction with a hitherto unknown field that modifies its 
effective mass with the energy.
At low energy, the interaction with the unknown field is 
weak, so that the apparent neutrino mass is in the $\eV$ range,
whereas at higher energies, the interaction becomes stronger
and leads to the observed~\cite{OPERA2011v2} large tachyonic masses.
We thus assume that the (bulk of the) neutrino mass is 
created dynamically~\cite{Oi2011}.
Possibly, there is some threshold region where the 
effective mass of the neutrino intersects with the 
mass of the field it interacts with, and this might help explain 
consistency with astrophysical data~\cite{DaEtAl1987}.
In the following, we would like to present a 
semi-quantitative analysis which supports these conjectures.

We investigate a scalar-minus-pseudoscalar
($S-P$) interaction Lagrangian of the form
\begin{equation}
\label{Lint}
\calL_{\rm int} = G \,\hat{\phi}_X \, 
\overline{\hat{\psi}} \, (1 - \gamma^5) \, \hat{\psi} \,.
\end{equation}
Here, $\hat{\phi}_X$ is a scalar field operator, $G$ is a dimensionless
coupling, and the fermionic field operators
for the neutrino are denoted as $\hat{\psi}$. 
The operator is of dimension~4 and therefore renormalizable;
it describes a Yukawa interaction with a chirality projector.
The complete Lagrangian of the tachyonic neutrino field,
the scalar field plus the $S-P$ interaction reads
\begin{align}
\label{LLLLL}
\calL(x) =& \; \frac{\ii}{2}
\left[ \overline{\hat \psi}(x) \gamma^\mu \left( \partial_\mu \hat{\psi}(x) \right) -
\left( \partial_\mu \overline{\hat \psi}(x) \right) \gamma^\mu \hat{\psi}(x) \right]
- \overline{\hat \psi}(x) \,\gamma^5 \, m_\nu \, \hat{\psi}(x)
\nonumber\\[2ex]
& \; - \frac12 \, \hat{\phi}_X(x) \, \left( \Box + M^2_X \right) \hat{\phi}_X(x)
+ G \,\hat{\phi}_X(x) \, \overline{\hat \psi}(x) \, (1 - \gamma^5) \, \hat{\psi}(x) \,.
\end{align}
At low energy, from dimensional analysis alone, 
the induced one-loop neutrino mass running via the renormalization group (RG)
can be written down as
\begin{equation}
\frac{\dd m_\nu}{\dd \ln(\mu)} = \mu \, \frac{\dd m_\nu}{\dd \mu} 
\propto \left[m_\nu(\mu)\right]^3 \; \left[G_X(\mu)\right]^2 \,,
\qquad
\qquad
\left[G_X(\mu)\right]^2 =
G^2_X \, \ln(\mu) =
\frac{G^2}{M_X^2} \, \ln(\mu) \,,
\end{equation}
where we assume a logarithmic running of the coupling
constant with the scale $\mu$.
Integrating the RG evolution equation,
\begin{equation}
\int \frac{\dd m_\nu}{m_\nu^3} = G_X^2 \, \int \frac{\dd\mu}{\mu} \,,
\qquad
\int\limits^{m_\nu(17 \, \GeV)}_{m_\nu(18\,\keV)} \frac{\dd m_\nu}{m_\nu^3} = 
G_X^2 \, \int^{17 \, \GeV}_{18\,\keV} \frac{\dd\mu}{\mu} \, \ln(\mu), 
\end{equation}
with $m_\nu(18 \, \keV) \approx 100 \, \eV$ (see Ref.~\cite{RoEtAl1991})
and $m_\nu(17 \, \GeV) \approx 117 \, \MeV$ (see Ref.~\cite{OPERA2011v2}),
we find that an $X$ particle of mass in the range $M_X \approx 1.4 \,\keV$
could potentially induce a neutrino mass running from about 100~eV
at 18~keV energies~\cite{RoEtAl1991} to 117~MeV at energies
of 17~GeV~\cite{OPERA2011v2}. Here, we assume that $G \approx 1$ and
a universal running
of the electron neutrino mass~\cite{RoEtAl1991} and the
muon neutrino mass~\cite{OPERA2011v2} with the energy.
The difference in the observed OPERA neutrino mass~\cite{OPERA2011v2} of $117 \, \MeV$
with low-energy neutrino data Refs.~\cite{RoEtAl1991,AsEtAl1994,AsEtAl1996,StDe1995,%
WeEtAl1999,LoEtAl1999,BeEtAl2008}, where masses in the $\eV$ range were
observed, suggests that significant neutrino mass running has to set in 
at energies much below $17\,\GeV$, so that we can safely assume that $M_X \ll 17\,\GeV$. 
This finding and the interaction~\eqref{Lint} is not described by any known 
particle in the standard model, and thus, 
our model constitutes a pertinent extension.
However, one may object that this treatment amounts to 
an application of a one-loop running of the mass in a 
domain which in view of $G \approx 1$ clearly is 
nonperturbative.

This high-energy limit could be analyzed as follows.
We first recall that in the
high-energy domain, where the effective neutrino mass
is in the $\MeV$ range (see Refs.~\cite{OPERA2011v2,KaEtAl1979})
we assume that the neutrino mass is (almost) exclusively
generated by the strong
(nonperturbative) self-interaction with the $X$ field.
It is interesting to observe that {\em polynomial}
behaviour of RG functions in the {\em strong-coupling domain}
has recently been obtained by a sophisticated
analysis of higher-order perturbative terms,
for the $\beta$ functions of $\phi^4$ theories
and of quantum electrodynamics~\cite{Su2001phi4,Su2001}.
If the mass of the $X$ particle is negligible as compared 
to the mass of the neutrino in the high-energy domain, 
then the mass scaling must be independent of $M_X$, and 
again, from dimensional analysis alone, we may conjecture that 
in the high-energy, strong-coupling limit,
\begin{equation}
\mu \, \frac{\dd m_\nu}{\dd \mu} 
\propto G^2 \, m_\nu  \,,
\qquad
\qquad
\int \frac{\dd m_\nu}{m_\nu} = 
K \, G^2\, \int \frac{\dd \mu}{\mu} \,,
\qquad
\qquad
m_\nu(\mu) = m_\nu(\mu_0) \,
\left( \frac{\mu}{\mu_0} \right)^{K \,G^2} \,,
\end{equation}
where $K$ is a constant of order unity.
In view of Eq.~\eqref{mmE}, if we assume that $G \approx 1/\sqrt{K}$, then
\begin{equation}
\label{sun}
m_\nu = m_\nu(E_\nu) 
= \eta \, \left( E_\nu \right)^{K \, G^2}
\approx \eta \, E_\nu \,,
\qquad
G \approx \frac{1}{\sqrt{K}} \,,
\qquad
\eta \approx \frac{1}{145} \,,
\end{equation}
where the value of $\eta$ is chosen such as to be 
consistent with Eq.~\eqref{mmE}.

\begin{figure}[t!]
\begin{minipage}[b]{0.45\linewidth}
\begin{center}
\includegraphics[height=0.6\linewidth]{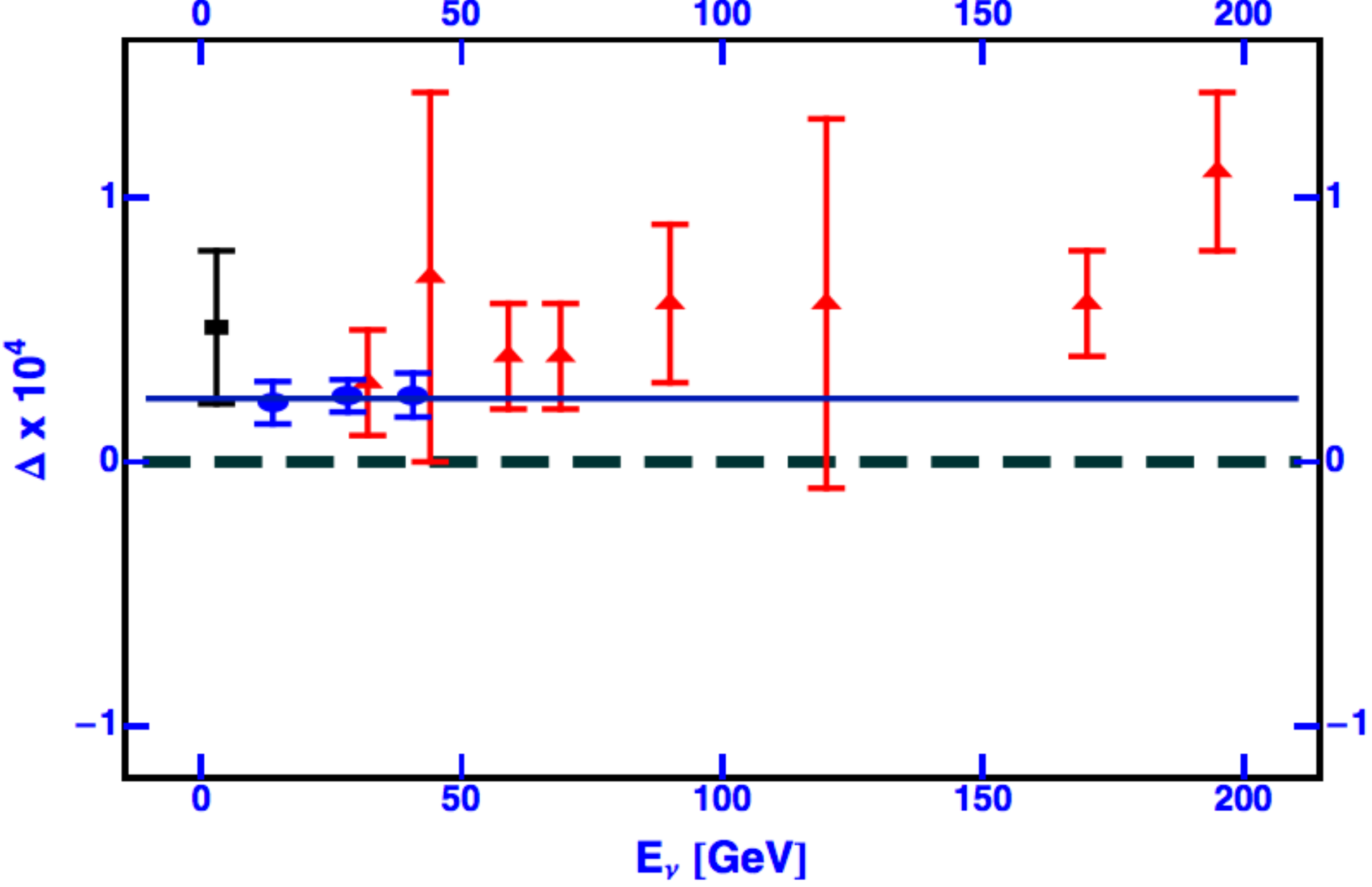} \\
{\bf (a)}
\end{center}
\end{minipage}
\begin{minipage}[b]{0.45\linewidth}
\begin{center}
\includegraphics[height=0.6\linewidth]{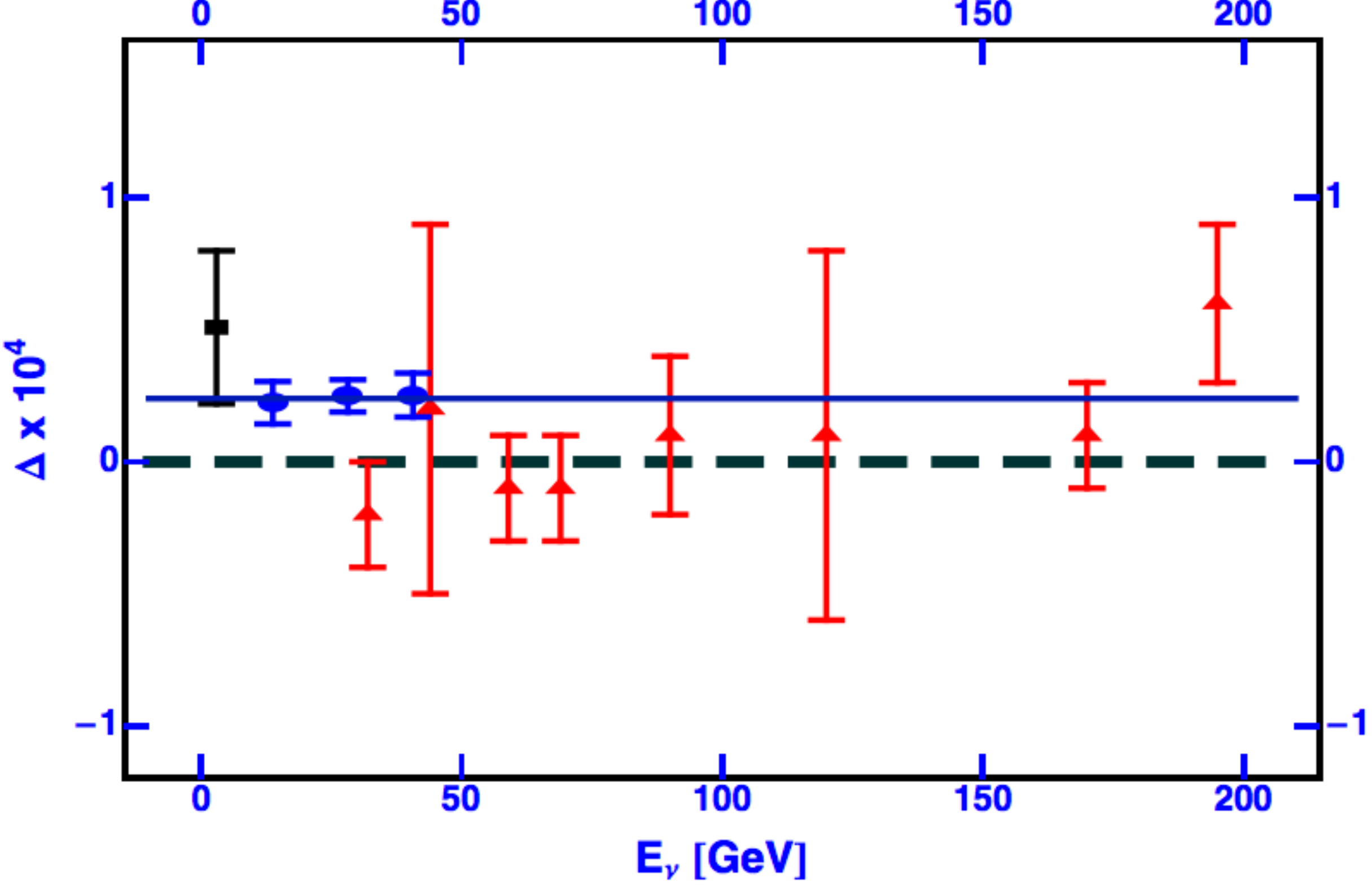} \\
{\bf (b)}
\end{center}
\end{minipage}
\caption{\label{fig3}(Color online.) Measured neutrino velocities in the 
range $E_\nu = 3 \, \GeV$ (Ref.~\cite{AdEtAl2007}) 
up to $E_\nu = 195 \, \GeV$ (Ref.~\cite{KaEtAl1979}).
The OPERA data are given in Eq.~\eqref{eee} 
and correspond to the data bins at
$E = 13.8 \, \GeV$, $E = 28.2 \, \GeV$, and $E = 40.7 \, \GeV$ (circles).
The data point at $E_\nu = 3 \, \GeV$ is from Ref.~\cite{AdEtAl2007} (square).
All remaining data points (triangles) are from Ref.~\cite{KaEtAl1979}.
Panel~(a) corresponds to the data plotted in Fig.~3 of Ref.~\cite{AdEtAl2007},
while panel~(b) applies a path length correction of 
$\Delta_{\rm path} = -0.5 \times 10^{-4}$ to the data (triangles) of 
Ref.~\cite{AdEtAl2007}, as discussed near the end of Ref.~\cite{AdEtAl2007}. 
Here, $\Delta$ is the relative 
deviation from the speed of light in vacuum,
which we multiply by a scaling factor $10^4$ on $y$ axis.
The solid line at $\Delta = 2.4 \times 10^{-4}$ 
corresponds to the result~\eqref{mres} based on our model.}
\end{figure}

We are now in the position to add some more,
somewhat speculative, remarks on the 
experimental findings of Refs.~\cite{KaEtAl1979,AdEtAl2007,OPERA2011v2}.
Based on the numerical entries in Table~II and Fig.~3 of Ref.~\cite{KaEtAl1979},
one may investigate the observed neutrino velocities as a function of the 
propagation energy.
The authors of the somewhat inconclusive
1979 paper (Ref.~\cite{KaEtAl1979})
suggest to ascribe a path length correction of 
$\Delta_{\rm path} = -0.5^{+0.2}_{-0.1} \times 10^{-4}$ 
to their data, because the muons that were ``racing'' against the 
neutrinos in the experiment were assumed to be artificially 
delayed due to multiple scattering events, which 
extend the muon path length in comparison to the 
muon neutrino path length. 
The path length correction was assumed to be constant over the 
energy range analyzed in Ref.~\cite{KaEtAl1979},
uniformly affecting neutrinos in the energy range of 
$32 \, \GeV < E_\nu < 195 \, \GeV$ in an experiment
over a relatively short baseline of about 900~m (which is 
smaller than the OPERA baseline by a factor of roughly $10^3$).
We find that the discussion on the derivation of the path length correction 
in Ref.~\cite{KaEtAl1979} is rather short and therefore 
present data with and without this correction in Fig.~\ref{fig3};
the same approach was recently taken in Figs.~1 and~2 of 
Ref.~\cite{ACEtAl2011a}.
The model~\eqref{sun} leads to a constant deviation of the 
neutrino velocity of 
\begin{equation}
\label{mres}
\Delta = \frac{v-c}{c} = \sqrt{1 + \eta^2} - 1 = 2.4 \times 10^{-5} \,,
\end{equation}
independent of the neutrino energy. This result is compared to the 
available experimental data~\cite{KaEtAl1979,AdEtAl2007,OPERA2011v2}
in Fig.~\ref{fig3}.
While our model is somewhat speculative at the current stage,
it is intriguing to observe that the solution of the 
simple-minded RG equation~\eqref{sun} is in good
agreement with the observed neutrino velocities over a wide 
energy interval (see Fig.~\ref{fig3}).
We also recall that the concomitant significant 
neutrino mass running will suppress decays because the
tachyonic mass in the exit channel is much lower than in the incoming channel
(see Sec.~\ref{kc}).

%
%
%
%
%

%
%
\section{Conclusions}
\label{conclu}

Tachyons have a potential of fundamentally altering our view of physical law,
but they can be incorporated into the framework of Lorentz transformations,
despite obvious problems with the causality principle.
In Ref.~\cite{ArSu1968}, the authors
argue that a ``sensible'' theory is obtained if one insists 
that the only physical quantities are transition amplitudes,
 and a negative-energy in
(out) state is understood to be a positive-energy out (in) state.
This statement is in need of further explanation.
Suppose that observer $A$ sees event $E$
before $E'$, and observer $A'$ sees event $E'$ before $E$, because the two
events are separated by a space-like interval, and the Lorentz transform
for the frames $A$ and $A'$ reverses the time-ordering of
events $E$ and $E'$. According to Ref.~\cite{BiDeSu1962}, the reversed
time ordering occurs if and only if 
the energy between the two frames also changes sign. So,
provided one reinterprets the negative-energy eigenstates of tachyonic Dirac
Hamiltonian propagating backward in time (the antiresonances included) as
positive-energy solutions propagating forward in time,
the creation and absorption of a particle can be consistently reinterpreted 
if only the transition amplitude
is unaffected by the reinterpretation. This point has also been stressed in
Refs.~\cite{BiDeSu1962,BiSu1969}.

One problem, though, in the consistency of observations of 
tachyons lies in conceivable decay processes~\cite{CoGl2011}.
In this paper, we investigate threshold conditions for the 
emission (see Sec.~\ref{kc}) of real particles by analogues of Cerenkov radiation 
emitted by superluminal, tachyonic neutrinos that fulfill
the dispersion relation~\eqref{dispsup}. We find that such 
emissions, as shown in Fig.~\ref{fig1},
are possible at high energies for small Cerenkov angles
in a narrow cone of emission angles $\theta$
[see Eqs.~\eqref{thetagamma} and~\eqref{theta0}]. Furthermore, at sufficiently 
large energy, a nonvanishing emission probability exists 
for even very small tachyonic mass squares $-m_\nu^2$. 
However, the calculation of the corresponding decay rates
crucially depends on the dispersion relation used in the calculation.
The tachyonic relation~\eqref{dispsup} is Lorentz-invariant, 
and the effective mass $m_\nu$ crucially influences the decay rate.
We then investigate, based on the tachyonic Dirac equation
(see Sec.~\ref{td}), how the effective neutrino mass $m_\nu$ could
possibly change from a few $\eV$ at low energies in the $\keV$
range to energies of a hundred $\MeV$ in the $\GeV$ range
[see Eqs.~\eqref{mmE1},~\eqref{mmE2} and~\eqref{mmE}].

We here come to the conclusion that 
a viable explanation for the large virtuality 
$E_\nu^2 - \vec p^{\,2}$ of the OPERA neutrinos 
could be due to an additional interaction that modifies the 
neutrino propagation at high energies. 
At an energy in the $\GeV$ range, as measured by OPERA,
the propagation velocity of
a particle with a rest mass on the order of a few $\eV$
is not expected to deviate from the speed of light by 
a factor on the order of $10^{-5}$. 
It does not really matter in this case that the OPERA experiment
has measured a deviation of $v_\nu$ from $c$ in the {\em super}luminal direction.
A hypothetical experimental result for $v_\nu - c < 0$ in the {\em sub}luminal 
direction, of the same order-of-magnitude,
as indicated in Eq.~\eqref{hypothetical}, would have been 
equally surprising. According to previous neutrino 
data~\cite{RoEtAl1991,AsEtAl1994,AsEtAl1996,StDe1995,%
WeEtAl1999,LoEtAl1999,BeEtAl2008},
OPERA was not expected to find a deviation $| v_\nu - c|$ in the 
neutrino propagation velocity of the order-of-magnitude given in
Eq.~\eqref{mmE}. In light of~Eqs.~\eqref{mmE1} and~\eqref{mmE2}, the OPERA
signal would otherwise 
correspond to a particle with a rest mass in the 
range of a hundred $\MeV$, or, with an effective 
mass of the neutrino that grows linearly with the energy.
Unfortunately, neither the 
Higgs mechanism nor the Gross-Neveu model,
induce a mass that depends on the energy.
Once the vacuum expectation value of the background field
that generates the mass is fixed, the mass of the 
constituent particle is also fixed. 
We find that it is indicated 
to investigate genuine neutrino mass running due to interactions
which have hitherto not been introduced into the standard model.
In Sec.~\ref{running} of this 
paper, we write down a chiral Yukawa interaction 
which might induce a neutrino mass running with the 
experimentally observed parameters.

%
%
\section*{Acknowledgments}

Helpful conversations with B.~J.~Wundt are gratefully acknowledged.
The author acknowledges support from the National Science Foundation
and by a Precision Measurement Grant from the National Institute of Standards
and Technology.


\begin{thebibliography}{10}

\bibitem{Re2009}
E. Recami, J. Phys. Conf. Ser. {\bf 196},  012020  (2009).

\bibitem{DaEtAl1987}
V.~L. Dadykin, G.~T. Zatsepin, V.~B. Karchagin, P.~V. Korchagin, S.~A. Mal'gin,
  O.~G. Ryazhskaya, V.~G. Ryasnyi, V.~P. Talochkin, F.~F. Khalchukov, V.~F.
  Yakushev, M. Aglietta, G. Badino, G. Bologna, C. Castagnoli, A. Castellina,
  W. Fulgione, P. Galeotti, O. Saavedra, J. Trincero, and S. Vernetto, JETP
  {\bf 45},  593  (1987).

\bibitem{AdEtAl2007}
P. Adamson {\it et~al.}, Phys. Rev. D {\bf 76},  072005  (2007).

\bibitem{KaEtAl1979}
G.~R. Kalbfleisch, N. Baggett, E.~C. Fowler, and J. Alspector, Phys. Rev. Lett.
  {\bf 43},  1361  (1979).

\bibitem{OPERA2011v2}
M. Antonello {\em et al.}, ICARUS Collaboration, {\em Measurement of the
  neutrino velocity with the ICARUS detector at the CNGS beam}, e-print
  arXiv:1203.3433v1.

\bibitem{RoEtAl1991}
R.~G.~H. Robertson, T.~J. Bowles, G.~J. Stephenson, D.~L. Wark, J.~F.
  Wilkerson, and D.~A. Knapp, Phys. Rev. Lett. {\bf 67},  957  (1991).

\bibitem{AsEtAl1994}
K. Assamagan, C. Br\"{o}nnimann, M. Daum, H. Forrer, R. Frosch, P. Gheno, R.
  Horisberger, M. Janousch, P.-R. Kettle, T. Spirig, and C. Wigger, Phys. Lett.
  B {\bf 335},  231  (1994).

\bibitem{StDe1995}
W. Stoeffl and D.~J. Decman, Phys. Rev. Lett. {\bf 75},  3237  (1995).

\bibitem{AsEtAl1996}
K. Assamagan, C. Br\"{o}nnimann, M. Daum, H. Forrer, R. Frosch, P. Gheno, R.
  Horisberger, M. Janousch, P.~R. Kettle, T. Spirig, and C. Wigger, Phys. Rev.
  D {\bf 53},  6065  (1996).

\bibitem{WeEtAl1999}
C. Weinheimer, B. Degen, A. Bleile, J. Bonn, L. Bornschein, O. Kazachenko, A.
  Kovalik, and E. Otten, Phys. Lett. B {\bf 460},  219  (1999).

\bibitem{LoEtAl1999}
V.~M. Lobashev, V.~N. Aseev, A.~I. Belesev, A.~I. Berlev, E.~V. Geraskin, A.~A.
  Golubev, O.~V. Kazachenko, Y.~E. Kuznetsov, R.~P. Ostroumov, L.~A. Ryvkis,
  B.~E. Stern, N.~A. Titov, S.~V. Zadorozhny, and Y.~I. Zakharov, Phys. Lett. B
  {\bf 460},  227  (1999).

\bibitem{BeEtAl2008}
A.~I. Belesev, E.~V. Geraskin, B.~L. Zhuikov, S.~V. Zadorozhny, O.~V.
  Kazachenko, V.~M. Kohanuk, N.~A. Lihovid, V.~M. Lobasheva, A.~A. Nozik, V.~I.
  Parfenov, A.~K. Skasyrskaya, E.~A. Sudachkov, N.~A. Titiov, and V.~G. Usanov,
  Phys. At. Nucl. {\bf 71},  449  (2008).

\bibitem{LABneutrino}
see the URL http://cupp.oulu.fi/neutrino/nd-mass.html.

\bibitem{ChHaKo1985}
A. Chodos, A.~I. Hauser, and V.~A. Kostelecky, Phys. Lett. B {\bf 150},  431
  (1985).

\bibitem{Ch2000}
T. Chang, {\em A new Dirac-type equation for tachyonic neutrinos}, e-print
  arXiv:hep-th/0011087.

\bibitem{Ch2002}
T. Chang, Nucl. Sci. Technol. {\bf 13},  129  (2002).

\bibitem{JeWu2011jpa}
U.~D. Jentschura and B.~J. Wundt, {\em Pseudo--Hermitian Quantum Dynamics of
  Tachyonic Spin--$\maybebm{1/2}$ Particles}, submitted, a preliminary version
  can be found as e-print arXiv:1110.4171.

\bibitem{So1905}
A. Sommerfeld, {\em Zur Elektronentheorie. III. Ueber Lichtgeschwindigkeits-
  und Ueberlichtgeschwindigkeitselektronen}, Nachr.~k.~Ges.~Wiss.
  G\"{o}ttingen, Heft 3, pp.~201-235 (1905).

\bibitem{Fe1967}
G. Feinberg, Phys. Rev. {\bf 159},  1089  (1967).

\bibitem{BiDeSu1962}
O.~M.~P. Bilaniuk, V.~K. Deshpande, and E.~C.~G. Sudarshan, Am. J. Phys. {\bf
  30},  718  (1962).

\bibitem{BaSh1974}
J. Bandukwala and D. Shay, Phys. Rev. D {\bf 9},  889  (1974).

\bibitem{vDNgBi1985}
H. van Dam, Y.~J. Ng, and L.~C. Biedenharn, Phys. Lett. B {\bf 158},  227
  (1985).

\bibitem{XiJi1987}
L. Xinzhou and L. Jizong, J. Phys. A {\bf 20},  6113  (1987).

\bibitem{Bi2009}
O.~M. Bilaniuk, J. Phys. Conf. Ser. {\bf 196},  012021  (2009).

\bibitem{Bo2009}
S.~K. Bose, J. Phys. Conf. Ser. {\bf 196},  012022  (2009).

\bibitem{DePe2003}
P. Defraigne and G. Petit, Metrologia {\bf 40},  184  (2003).

\bibitem{Le2008}
J. Levine, Metrologia {\bf 45},  184  (2008).

\bibitem{CoGl2011}
A.~G. Cohen and S.~L. Glashow, Phys. Rev. Lett. {\bf 107},  181803  (2011).

\bibitem{AC2000}
G. Amelino-Camelia, {\em Relativity in space-times with short-distance
  structure governed by an observer-independent (Planckian) length scale},
  e-print arXiv:gr-qc/0012051.

\bibitem{AC2010}
G. Amelino-Camelia, Symmetry {\bf 2},  230  (2010).

\bibitem{ACEtAl2011a}
G. Amelino-Camelia, G. Gubitosi, N. Loret, F. Mercati, G. Rosati, and P.
  Lipari, {\em OPERA-reassessing data on the energy dependence of the speed of
  neutrinos}, arXiv:1109.5172.

\bibitem{ACEtAl2011b}
G. Amelino-Camelia, L. Freidel, J. Kowalski-Glikman, and L. Smolin, {\em OPERA
  neutrinos and deformed special relativity}, arXiv:1110.0521.

\bibitem{BiYiYuYu2011}
X.-J. Bi, P.-F. Yin, Z.-H. Yu, and Q. Yuan, Phys. Rev. Lett. {\bf 107},  241802
   (2011).

\bibitem{CoNuSa2011}
R. Cowsik, S. Nussinov, and U. Sarkar, Phys. Rev. Lett. {\bf 107},  251801
  (2011).

\bibitem{GM2011}
L. Gonzalez-Mestres, {\em Astrophysical consequences of the OPERA superluminal
  neutrino}, arXiv:1109.6630.

\bibitem{ArSu1968}
M.~E. Arons and E.~C.~G. Saudarshan, Phys. Rev. {\bf 173},  1622  (1968).

\bibitem{DhSu1968}
J. Dhar and E.~C.~G. Saudarshan, Phys. Rev. {\bf 174},  1808  (1968).

\bibitem{SuSh1986}
R.~I. Sutherland and J.~R. Shepanski, Phys. Rev. D {\bf 33},  2896  (1986).

\bibitem{MoRa2011}
S. Mohanty and S. Rao, {\em Constraint on super-luminal neutrinos from vacuum
  Cerenkov processes}, arXiv:1111.2725.

\bibitem{LiLiMeWaZh2011}
M. Li, D. Liu, J. Meng, T. Wang, and L. Zhou, {\em Replaying neutrino
  bremsstrahlung with general dispersion relations}, arXiv:1111.3294.

\bibitem{Je2001pra}
U.~D. Jentschura, Phys. Rev. A {\bf 64},  013403  (2001).

\bibitem{Pa1943}
W. Pauli, Rev. Mod. Phys. {\bf 15},  175  (1943).

\bibitem{BeBo1998}
C.~M. Bender and S. Boettcher, Phys. Rev. Lett. {\bf 80},  5243  (1998).

\bibitem{BeDu1999}
C.~M. Bender and G.~V. Dunne, J. Math. Phys. {\bf 40},  4616  (1999).

\bibitem{BeBoMe1999}
C.~M. Bender, S. Boettcher, and P.~N. Meisinger, J. Math. Phys. {\bf 40},  2201
   (1999).

\bibitem{BeBrJo2002}
C.~M. Bender, D.~C. Brody, and H.~F. Jones, Phys. Rev. Lett. {\bf 89},  270401
  (2002).

\bibitem{Mo2002i}
A. Mostafazadeh, J. Math. Phys. {\bf 43},  205  (2002).

\bibitem{Mo2002ii}
A. Mostafazadeh, J. Math. Phys. {\bf 43},  2814  (2002).

\bibitem{Mo2002iii}
A. Mostafazadeh, J. Math. Phys. {\bf 43},  3944  (2002).

\bibitem{Mo2003npb}
A. Mostafazadeh, J. Math. Phys. {\bf 44},  974  (2003).

\bibitem{Bo1970}
D.~G. Boulware, Phys. Rev. D {\bf 1},  2426  (1970).

\bibitem{MoiseyevMcCurdy}
N. Moiseyev, Phys. Rep. {\bf 302}, 211 (1998); C. W. McCurdy and C. K. Stroud,
  Comput. Phys. Commun. {\bf 63}, 323 (1991).

\bibitem{JeSuLuZJ2008}
U.~D. Jentschura, A. Surzhykov, M. Lubasch, and J. Zinn-Justin, J. Phys. A {\bf
  41},  095302  (2008).

\bibitem{ItZu1980}
C. Itzykson and J.~B. Zuber, {\em \relax{Quantum Field Theory}} (McGraw-Hill,
  New York, 1980).

\bibitem{JeWu2012}
U.~D. Jentschura and B.~J. Wundt, Eur. Phys. J. C {\bf 72},  1894  (2012).

\bibitem{Oi2011}
V.~K. Oikonomou, {\em The 2d Gross--Neveu Model for Pseudovector Fermions and
  Tachyonic Mass Generation}, e-print arXiv:1109.6170.

\bibitem{Su2001phi4}
I.~M. Suslov, Zh. \'{E}ksp. Teor. Fiz. {\bf 120},  5  (2001), [JETP {\bf 93}, 1
  (2001)].

\bibitem{Su2001}
I.~M. Suslov, Pis'ma v ZhETF {\bf 74},  211  (2001), [JETP Lett.~{\bf 74}, 191
  (2001)].

\bibitem{BiSu1969}
O.-M. Bilaniuk and E.~C.~G. Sudarshan, Nature (London) {\bf 223},  386  (1969).

\end{thebibliography}
\end{document}